**Predicting influenza H3N2 vaccine efficacy from evolution of the dominant epitope**

Melia E. Bonomo, Department of Physics and Astronomy and Center for Theoretical Biological Physics, Rice University, Houston, TX, USA

Michael W. Deem, Department of Bioengineering, Department of Physics and Astronomy and Center for Theoretical Biological Physics, Rice University, Houston, TX, USA

**Running title:** Predicting influenza vaccine efficacy

**40-word summary of main point**: Our $p_{\text{epitope}}$ model predicts the ability of the influenza vaccine to reduce the A(H3N2) disease attack rate, with an r^2=0.77. This fast, sequence-based method compliments strain-to-strain antigenic comparisons from ferret models and provides antigenic comparisons for all circulating sequences.

**Corresponding author:** Michael W. Deem, 6500 Main St, STE 1030, Rice University, Houston, TX 77030, USA; mwdeem@rice.edu, +17133485852.

**Alternate corresponding author:** Melia E. Bonomo, Dept. of Physics and Astronomy MS-61, Rice University, PO Box 1892, Houston, TX 77251, USA; meb16@rice.edu, +17133482204.

**Abstract:**

We predict vaccine efficacy with a measure of antigenic distance between influenza A(H3N2) and candidate vaccine viruses based on amino acid substitutions in the dominant epitopes. In 2016-2017, our model predicts 19% efficacy compared to 20% observed. This tool assists candidate vaccine selection by predicting human protection against circulating strains.

**Text:**

Seasonal influenza constitutes a significant disease burden worldwide, with three to five million cases of severe illness and an estimated annual death toll of 290,000 to 650,000; however, vaccination can provide protection [1]. The vaccine component against influenza A(H3N2) is especially important, as increased morbidity and mortality are associated with this most commonly predominant subtype [2]. Influenza type A viruses are primarily recognized by the immune system via two proteins on their surface, hemagglutinin (HA) and neuraminidase [3]. These viruses constantly evolve to evade human antibodies, most notably by introducing amino acid substitutions into the HA binding sites, designated epitopes *A* through *E* in A(H3N2) (Supplementary Figure 1). Some epitopes appear to play a more dominant role during infection than others, and the one under the greatest immune pressure in a given season will have the highest percentage of amino acid substitutions, computed by dividing the number of substitutions by the total number of amino acids in that epitope [4]. Increased antigenic distance between the vaccine and infecting virus leads to decreased vaccine efficacy. Due to virus evolution, the World Health Organization (WHO) recommends the composition of the seasonal influenza vaccine twice a year based on the dominant influenza strains from the previous northern or southern hemisphere season [5]. Several antigenically similar candidate vaccine viruses (CVVs) and associated reassortants are made available. Generally, the reassorted viruses used in manufacturing the vaccine are antigenically identical to their CVV prototypes in the HA region [6]. For the 2016-2017 northern hemisphere influenza season, WHO recommended A/Hong Kong/4801/2014 (H3N2)-like CVVs [5]. However, vaccine efficacy for adults aged 18-64 against A(H3N2) influenza was only 20 ± 8% [2] (Supplementary Methods). Rather than an increased antigenic distance due to virus evolution, the 2016-2017 vaccine may have diverged from circulating viruses due to substitutions acquired during isolation of the CVV strains in eggs [7]. Passaging-related adaptations have posed an issue for A(H3N2) CVVs in particular [8,9] (Supplementary Table 3).

We have previously derived a statistical mechanics model that captures the dynamics of human antibody-mediated response to viral infection following vaccination [10]. In a recent application of our theory, we used the evolution of the dominant HA epitope to predict vaccine efficacy, the ability of the vaccine to reduce the disease attack rate [4]. Here we generalize the model to A(H3N2) data both from the 1971-1972 to 2015-2016 influenza seasons and from highly consistent, laboratory-confirmed studies over the past decade. Additionally, we expand the prediction to encompass average efficacy against an abundance of diverse A(H3N2) strains for a given season. We apply this novel approach to quantify antigenic distance and expected efficacy of the egg-adapted CVVs against all strains circulating during the 2016-2017 and early 2017-2018 seasons.

**Methods**

The epitope-based dependence of vaccine efficacy in our model comes from a measure of antigenic distance termed $p_{\text{epitope}}$, where

$$p_{\text{epitope}} = \frac{\text{number of substitutions in dominant epitope}}{\text{total number of amino acids in dominant epitope}} \qquad (1)$$

In ferret models, the typical measure of antigenic distance $d_1$ is the $\log_2$ difference between vaccine antiserum titer against itself and the vaccine antiserum titer against a strain representative of the dominant circulating viruses [11]. A second measure $d_2$ is the square-root ratio of the product of the homologous titers to the product of the heterologous titers [12]. We perform linear regression with vaccine efficacy data reported by WHO collaborating centers to compare the predictive power of $p_{\text{epitope}}$ with these ferret-based distances. We then utilize our model to evaluate the primary protein structure of egg-adapted A/Hong Kong/4801/2014 CVV in the context of the 2016-2017 season [5] and to predict efficacy of the egg-adapted CVV for the newly recommended A/Singapore/INFIMH-16-0019/2016 (H3N2)-like vaccine [13]. Finally, we calculate the average efficacy of the CVVs against 6610 circulating A(H3N2) strains collected from September 2016 to November 2017, a unique capability of our $p_{\text{epitope}}$ measure of antigenic distance in comparison to other clinical studies.

**Results**

Our use of $p_{\text{epitope}}$ to correlate antigenic distance with vaccine efficacy yields a coefficient of determination $r^2 = 0.77$ on both data since 1971 and data over just the past 10 years (Figure 1). Conventional ferret $d_1$ has $r^2 = 0.42$ on data since 1971 and has dropped to $r^2 = 0.23$ on data over the past 10 years. The extracted equation to predict vaccine efficacy $VE$ is

$$VE = -2.471 p_{\text{epitope}} + 0.468,$$

(2)

where the standard error on the slope is $\pm0.254$ and the standard error on the y-intercept is $\pm0.032$ (Supplementary Methods).

We calculate $p_{\text{epitope}} = 0.095$ between the egg-adapted A/Hong Kong CVV and A/Colorado/15/2014, a representative clade 3C.2a strain [7], and find the dominant epitope is $B$, with amino acid substitutions T160K and L194P. The reassorted CVVs contain some additional substitutions (Supplementary Figure 2), but as these substitutions are outside of the dominant epitope, we focus on the non-reassorted CVV for our analysis. Using Equation 2, we predict vaccine efficacy against this A/Colorado strain to be 23 ± 4% (Supplementary Table 1). The hemagglutination inhibition titers from post-infection ferret antisera reported for the 2016-2017 season [14] and our linear regressions of $d_1$ and $d_2$ predict vaccine efficacy to be 27 ± 5% and 28 ± 4%, respectively. Analysis of the reference A/Hong Kong strain, which was the clinical specimen collected from humans and expected to dominate in the population, reveals that this strain did not exhibit the epitope $B$ substitutions. We predict it would have had a high 47 ± 3% efficacy against A/Colorado.

When comparing the CVVs to all circulating strains during the 2016-2017 and early 2017-2018 seasons, we considered both $p_{epitope}$ (Equation 1) and the Hamming distance, which is the number of differing amino acids between each pair of strains in HA subunit 1 divided by the total 328 amino acids of the subunit. While the Hamming distance between the reference A/Hong Kong specimen and the consensus sequence of the circulating strains is 0.006, that of the egg-adapted CVV's is 0.015 (Supplementary Figure 2). For the egg-adapted A/Hong Kong CVV, the average $p_{epitope}$ is 0.111 due to substitutions in dominant epitope $B$, with an average predicted vaccine efficacy of 19 ± 4%. The reference A/Singapore/INFIMH-16-0019/2016 specimen has a Hamming distance of 0.003 to the consensus strain, smaller than the reference A/Hong Kong specimen has, and that of its egg-adapted CVV is 0.012. We calculate that the egg-passaged A/Singapore CVV has an average $p_{epitope}$ of 0.118, leading to a predicted average vaccine efficacy of 18 ± 4%.

**Discussion**

We quantify influenza virus evolution and predict the ability of the CVV to reduce the disease attack rate using our $p_{epitope}$ measure of antigenic distance. While other factors influence real-world vaccine effectiveness in humans, $p_{epitope}$ has accounted for much of the variance in CVV efficacy over the past 10 and 45 years ($r^2 = 0.77$). The low vaccine performance against A(H3N2) during 2016-2017 appears to have been caused by the T160K and L194P substitutions that occurred in the dominant epitope *B* during egg-isolation of the A/Hong Kong CVV. Indeed, human antisera studies showed that individuals who had received the Flublok recombinant influenza vaccine (containing T160 and L194) that had been passaged in insect cells produced more effective antisera than did those who had received the inactivated influenza vaccines Fluzone or Flucelvax (containing K160 and P194) [7], which were both made from CVVs initially isolated in eggs [9]. Arguably, HA stability did not account for the larger response induced by Flublok, since all three vaccines induced a similar antisera response when challenged by K160-containing viruses [7].

It is critical to choose a vaccine strain with the minimal antigenic distance from all circulating A(H3N2) strains. While the reference A/Singapore specimen minimizes the Hamming distance, the reference A/Hong Kong specimen minimizes $p_{epitope}$ (Supplementary Figure 3). WHO changed the recommended CVVs for the southern hemisphere because the egg-passaged A/Singapore CVV contains an N121K substitution that matches the majority of recent viruses, and ferret antisera were more successfully raised [13]. However, the egg-passaged A/Singapore CVV has two substitutions in residue *B* (T160K and L194P) that do not match the majority of circulating strains. We predict that this A/Singapore CVV will have comparable efficacy to the egg-passaged A/Hong Kong CVVs in humans.

The $p_{\text{epitope}}$ of a pair of strains can be calculated nearly instantaneously or averaged for one strain against 6000 in seconds. By contrast, ferret models are restricted to a few analysis pairs, antisera production takes 3-5 weeks, and there is added difficultly due to strict biocontainment measures [9]. Additionally, $p_{\text{epitope}}$ theory suggests which sequence discrepancies are contributing most to a lowered vaccine efficacy. We currently do not explicitly consider N-linked glycosylation, though $p_{\text{epitope}}$ implicitly incorporates this seeing as discrepancies in glycosylation between two strains are generally due to amino acid substitutions, *e.g.,* T160K caused an observed lack of glycosylation site in the A/Hong Kong CVV as compared to circulating 3C.2a strains [7]. Our model can rapidly narrow down clinical specimens that are representative of viruses that will dominate the upcoming flu season. It can then be used in cooperation with ferret models to confirm that the CVVs and their associated reassortants have not acquired critical antigenic changes before the vaccine is manufactured and administered to the human population worldwide.

**Funding**

This work was supported by the Center for Theoretical Biological Physics at Rice University, Houston TX [National Science Foundation, PHY 1427654] and the Welch Foundation [C--1917--20170325].

**Conflict of Interest**

The authors declare that they have no competing financial interests.

**Acknowledgments**

Influenza strain sequences were retrieved from the open access EpiFlu database hosted by GISAID: Global initiative on sharing all influenza data, accessible at http://platform.gisaid.org/. See Supplementary Table 2 for specific contributions of both the submitting and the originating laboratories.

**Figure 1.**

Influenza A(H3N2) vaccine efficacy as a function of three measures of antigenic distance. We plot $p_{\text{epitope}}$ (red) from our model, $d_1$ (green) and $d_2$ (blue) from ferret serological studies [11,12], and the corresponding vaccine efficacy values for each influenza season. Linear regression of data from each measure of antigenic distance is shown for the past 45 years (solid lines) and for laboratory-confirmed studies in the past 10 years (dotted lines). The coefficients of determination ($r^2$) are listed in the legend. Data for the 2016-2017 season are plotted separately (diamond points). The strong and consistent $r^2$ between $p_{\text{epitope}}$ and vaccine efficacy indicates that this model accurately predicts the data. We determined an average efficacy of $19 \pm 4\%$ for the 2016-2017 egg-adapted A/Hong Kong/4801/2014 vaccine, whereas $20 \pm 8\%$ was observed [2].

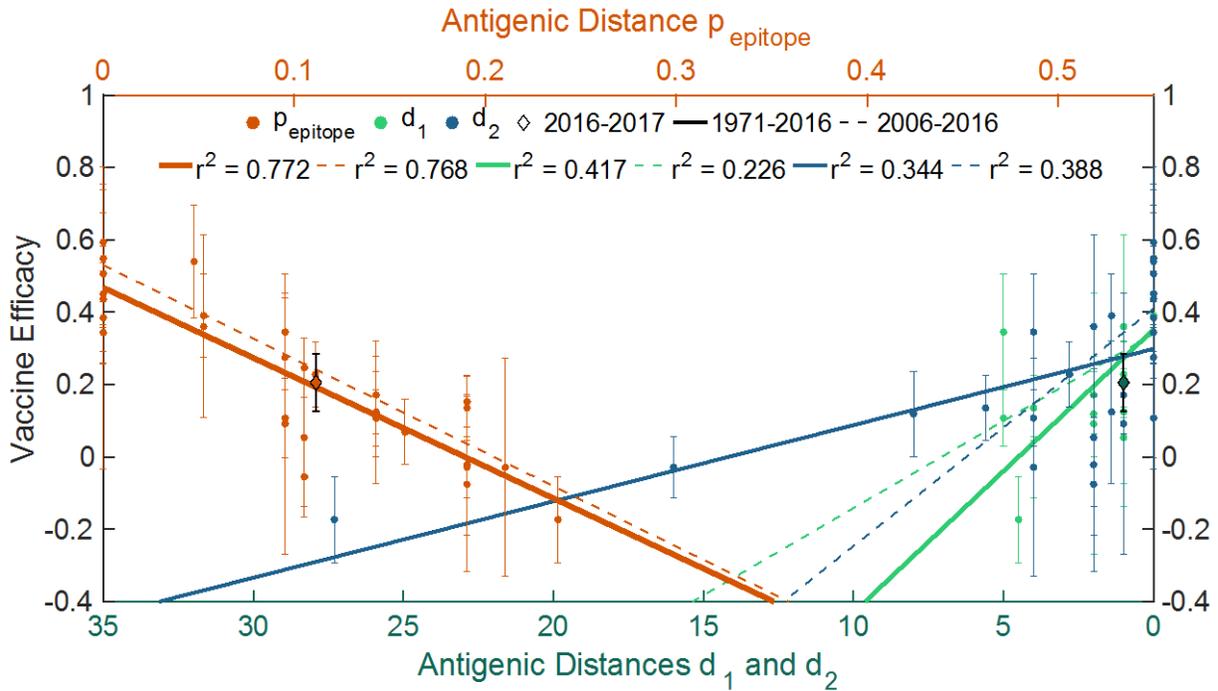

**Supplemental Methods**

<u>Calculation of $p_{epitope}$ and Hamming distance from sequence data.</u>

All hemagglutinin (HA) sequence data were obtained from the GISAID Platform's open access EpiFlu database [1] (Supplementary Table 2). We downloaded all sequences of H3N2 clinical specimens collected from humans between September 2016 and November 2017 that had been uploaded before 18 November 2017 (totaling 6610 HA sequences). To determine the dominant epitope and $p_{epitope}$, we select the 328 amino acids that are part of subunit one of the HA protein (HA1) and calculate the amino acid discrepancies for individual epitopes, scaled by the total number of amino acids in each epitope (Equation 1; Supplementary Figure 2). The consensus strain is the sequence containing the most common amino acid in each position among all sequences. The Hamming distance between each pair of strains is the number of amino acids discrepancies in the complete HA1 region, divided by the size of HA1 (328 amino acids),

$$d_{\text{Hamming}} = \frac{\text{number of substitutions in HA1}}{\text{total number of amino acids in HA1}} \qquad (S1).$$

<u>Calculation of antigenic distance from ferret serological studies.</u>

We use hemagglutination inhibition assay results to obtain the vaccine antiserum titer against itself $H_{11}$, the vaccine antiserum titer against a strain representative of the dominant circulating viruses $H_{21}$, the dominant strain antiserum titer against itself $H_{22}$, and the dominant strain antiserum titer against the vaccine strain $H_{12}$. The equations for antigenic distances $d_1$ [2] and $d_2$ [3] are defined as,

$$d_1 = \log_2 \frac{H_{11}}{H_{21}} \quad \text{and} \quad d_2 = \sqrt{\frac{H_{11}H_{22}}{H_{21}H_{12}}}. \quad (S2)$$

Vaccine efficacy.

We define vaccine efficacy from the rate of infection in unvaccinated individuals $u = n_u/N_u$ and the rate of infection in vaccinated individuals $v = n_v/N_v$ as

$$VE = \frac{u - v}{u}, \quad (S3)$$

where $N_u$ is the total number of unvaccinated individuals, $N_v$ is the total number of vaccinated individuals, $n_u$ is the number of unvaccinated individuals who tested H3N2-positive, and $n_v$ is the number of vaccinated individuals who tested H3N2-positive. Estimates of the standard error for these rates and vaccine efficacy are

$$\sigma_u = \sqrt{\frac{u(1-u)}{N_u}}, \quad \sigma_v = \sqrt{\frac{v(1-v)}{N_v}}, \text{ and } \sigma_{VE} = \frac{v}{u}\sqrt{\frac{{\sigma_v}^2}{v} + \frac{{\sigma_u}^2}{u}}. \quad (S4)$$

To minimize inevitable bias in vaccine efficacy due to individuals with a non-optimal immune system or differences in epidemiological study site procedures, we select only healthy adults aged 18-64 years and individuals vaccinated by inactivated viruses. We use the data for our population of interest from [4]. For the combined 18-64 cohort, 322 individuals were A(H3N2) positive, 143 of which had been vaccinated, meaning 179 had not been vaccinated. Additionally, the data shows that 1208 individuals were A(H3N2) negative, 624 of which had been vaccinated, meaning 584 had not been vaccinated. From these numbers, we see that there were 763 total unvaccinated individuals and 767 total vaccinated individuals. It then follows that the rate of infection (Supplementary Equation S3) in unvaccinated individuals is 179 / 763 = 0.2346 and the rate of infection in vaccinated individuals is 143 / 767 = 0.1864. The A(H3N2) vaccine efficacy for this cohort is then (0.235 - 0.049) / 0.235 = 0.2053, or 20%. We then calculated the standard error to be $\pm$ 8% using Supplementary Equation S4.

The standard error for the vaccine efficacy predicted by each measure of antigenic distance through linear regression is

$$\sigma_{VE} = \sqrt{\sigma_m^2 x^2 + \sigma_b^2},$$ (S5)

where $\sigma_m$ is the standard error on the slope, $\sigma_b$ is the standard error on the y-intercept, and $x$ is either $p_{epitope}$, $d_1$, or $d_2$.

**Supplementary Figure 1**

Protein structure of HA for an H3N2 strain.  Amino acid residues in subunit one of HA are filled and color-coded for each epitope, A (red, 19 amino acids), B (yellow, 21 amino acids), C (green, 27 amino acids), D (blue, 41 amino acids), and E (purple, 22 amino acids) [5,6].  The two dominant-epitope amino acid residues differing between clade 3C.2a circulating viruses and the egg-adapted candidate vaccine viruses (CVVs) for A/Hong Kong/4801/2014 and A/Singapore/INFIMH-16-0019/2016 are colored darker gold and labeled.

Figure generated with MATLAB Molecule Viewer using Protein Data Bank entry 2VIU [7].

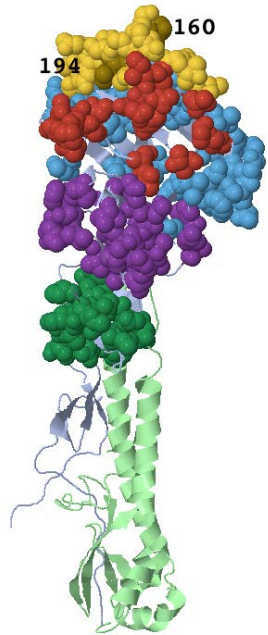

**Supplementary Figure 2**

The fractional number of substitutions in each HA epitope, as determined by dividing the number of substitutions by the total number of amino acids in the epitope. Clinical specimen reference (Ref.) strains, A/Hong Kong/4801/2014 and A/Singapore/INFIMH-16-0019/2016, and egg-adapted (Egg) CVVs are compared to the dominant circulating viruses, (**A**) A/Colorado/15/2014 and (**B**) A/Singapore/INFIMH-16-0019/2016. Discrepancies are detailed inside the bar segments, e.g., T160K means there is a T to K substitution in residue 160. The $p_{\text{epitope}}$ is calculated according to Equation 1 from the dominant epitope, which is the epitope that has the largest fractional discrepancy. The reassorted CVVs for A/Hong Kong (X263, X263A, and X263B) and for A/Singapore (IVR-186, NIB-104, X-307, and X-307A) are also compared to (**C**) A/Hong Kong (Egg) and (**D**) A/Singapore (Egg), respectively.

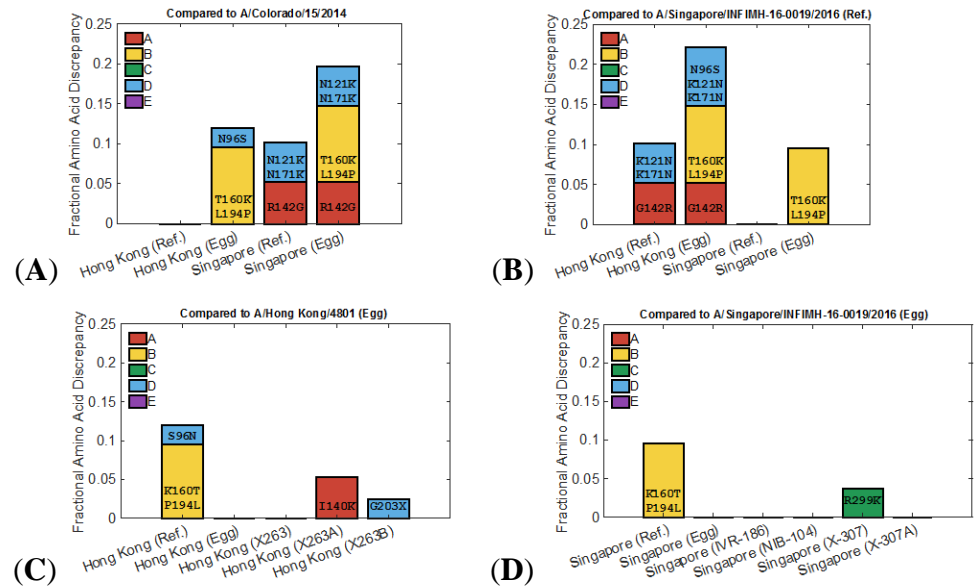

**Supplementary Figure 3**

We performed dimensional reduction using multidimensional scaling of the (**A**) Hamming distance and (**B**) $p_{epitope}$ between all circulating strains to display the comparison of sequence-space minimization of the clinical specimen reference and egg-adapted CVV strains. These plots contain 6339 H3N2 strains collected from September 2016 to August 2017 and 271 strains collected from September 2017 to November 2017. There does not seem to be a significant shift in the H3N2 quasispecies cluster at this point in the 2017-2018 season. Surveillance of the rates of infection and dominant circulating strains over the remainder of this 2017-2018 influenza season will reveal whether a new, distinct cluster is forming.

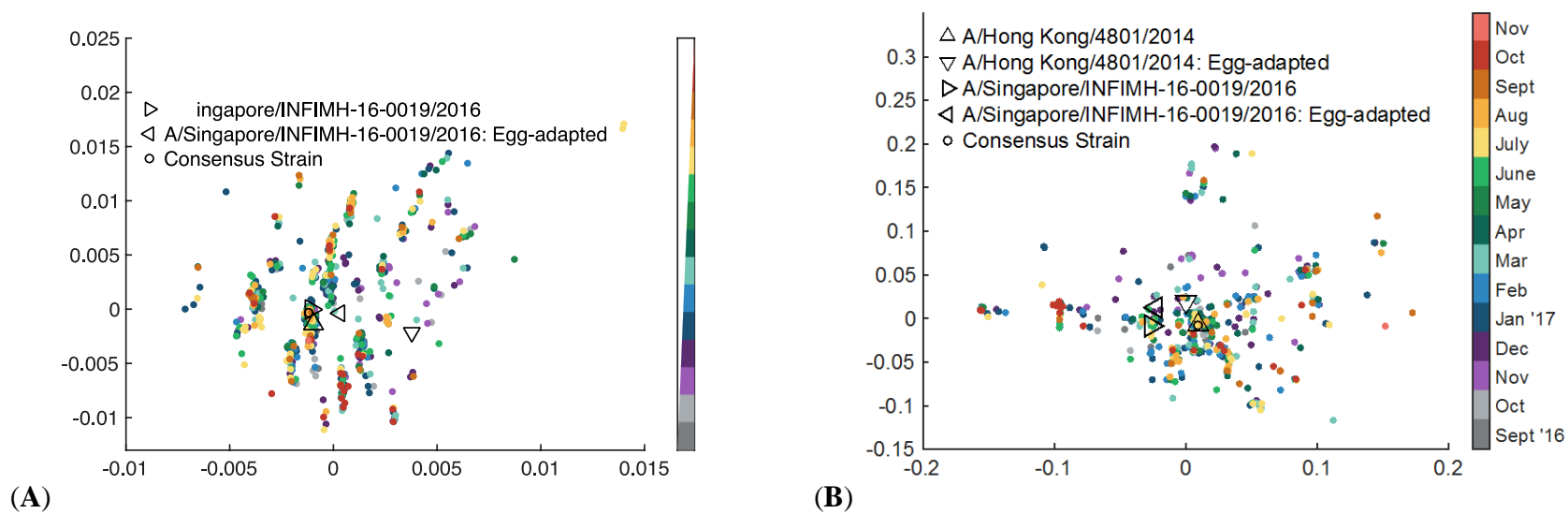

**Supplementary Table 1**

Summary of strains analyzed with our model. The value of $p_{\text{epitope}}$ is calculated with Equation 1, predicted vaccine efficacy is calculated with Equation 2 and Supplementary Equation S5, and observed vaccine efficacy it is calculated with Supplementary Equations S3 and S4 ($n_u$, $N_u$, $n_v$, $N_v$ are defined with Supplementary Equation S3). Serological measures of antigenic distance ($d_1$, $d_2$) are calculated from Supplementary Equation S2. The accession numbers for the GISAID database are listed under each isolate.

| Season | Vaccine | Circulating Strain | Dominant epitope | $p_{\text{epitope}}$ | Predicted Vaccine Efficacy with $p_{\text{epitope}}$ | Observed Vaccine Efficacy | $n_u$ | $N_u$ | $n_v$ | $N_v$ | $d_1$ | $d_2$ |
|---|---|---|---|---|---|---|---|---|---|---|---|---|
| 2015-16 | A/Switzerland/9715293/2013 - Egg<br>EPI_ISL_164719 | A/Hong Kong/4801/2014<br>EPI_ISL_165554 | A | 0.2105 | -5.5 ± 6.5 % | -2.9 ± 30.1 %<br>[8] | 27 | 1699 | 20 | 1223 | 4 | 4<br>[9] |
| 2016-17 | A/Hong Kong/4801/2014<br>EPI_ISL_165554 | A/Colorado/15/2014<br>EPI_ISL_167408 | none | 0 | 46.8 ± 3.2% | 20 ± 7.9 %<br>[4] | 179 | 763 | 143 | 767 | 1 | 1<br>[10] |
| 2016-17 | A/Hong Kong/4801/2014 - Egg<br>EPI_ISL_189814 | A/Colorado/15/2014<br>EPI_ISL_167408 | B | 0.0952 | 23.3 ± 4.0% | | | | | | | |
| 2016-17 | A/Hong Kong/4801/2014 - Egg<br>EPI_ISL_189814 | A/Hong Kong/4801/2014<br>EPI_ISL_165554 | B | 0.0952 | 23.3 ± 4.0% | | | | | | | |
| 2016-17 and early 2017-18 | A/Hong Kong/4801/2014 - Egg<br>EPI_ISL_189814 | All Circulating Strains | B | 0.1114 | 19.3 ± 4.3% | | | | | | | |
| 2016-17 and early 2017-18 | A/Hong Kong/4801/2014<br>EPI_ISL_165554 | All Circulating Strains | A | 0.078 | 27.5 ± 3.8% | | | | | | | |
| 2016-17 and early 2017-18 | A/Singapore/INFIMH-16-0019/2016<br>EPI_ISL_225834 | All Circulating Strains | A | 0.0915 | 24.2 ± 4.0% | | | | | | | |
| 2016-17 and early 2017-18 | A/Singapore/INFIMH-16-0019/2016 - Egg<br>EPI_ISL_239803 | All Circulating Strains | B | 0.1175 | 17.8 ± 4.4% | | | | | | | |
| 2018 | A/Singapore/INFIMH-16-0019/2016 - Egg<br>EPI_ISL_239803 | A/Singapore/INFIMH-16-0019/2016<br>EPI_ISL_225834 | B | 0.0952 | 23.3 ± 4.0% | | | | | | | |

**Supplementary Table 2**

GISAID Acknowledgements. We gratefully acknowledge the originating and submitting laboratories of the hemagglutinin (HA) sequences from GISAID's EpiFlu[TM] Database on which this research is based [1]. The list is detailed below. All submitters of data may be contacted directly via the GISAID website (http://www.gisaid.org).

| Isolate ID | Segment ID | Segment | Country | Date | Isolate Name | Originating Laboratory | Submitting Laboratory |
|---|---|---|---|---|---|---|---|
| EPI_ISL_164719 | EPI537866 | HA | Switzerland | 2013-Dec-6 | A/Switzerland/9715293/2013 | National Institute for Medical Research | Centers for Disease Control and Prevention |
| EPI_ISL_165554 | EPI539576 | HA | Hong Kong | 2014-Feb-26 | A/Hong Kong/4801/2014 | Government Virus Unit | National Institute for Medical Research |
| EPI_ISL_218060 | EPI741474 | HA | UK | 2016-Apr-22 | A/Hong Kong/4801/2014 (15/192) | Francis Crick Institute | National Institute for Biological Standards and Control |
| EPI_ISL_189810 | EPI614406 | HA | USA | 2014 | A/Hong Kong/4801/2014 X-263 | New York Medical College | Centers for Disease Control and Prevention |
| EPI_ISL_215973 | EPI731467 | HA | UK | 2016-Apr-5 | A/Hong Kong/4801/2014 X263A (15/188) | New York Medical College | National Institute for Biological Standards and Control |
| EPI_ISL_215974 | EPI731469 | HA | UK | 2016-Apr-5 | A/Hong Kong/4801/2014 X263B (15/184) | New York Medical College | National Institute for Biological Standards and Control |
| EPI_ISL_167408 | EPI545672 | HA | USA | 2014-Oct-4 | A/Colorado/15/2014 | Colorado Department of Health Laboratory | Centers for Disease Control and Prevention |
| EPI_ISL_225834 | EPI780183 | HA | Singapore | 2016-Jun-14 | A/Singapore/INFIMH-16-0019/2016 | Ministry of Health, Singapore | Ministry of Health, Singapore |
| EPI_ISL_282899 | EPI1086288 | HA | UK | 2017-Oct-24 | A/Singapore/INFIMH-16-0019/2016 (17/196) | Victorian Infectious Diseases Reference Laboratory | National Institute for Biological Standards and Control |
| EPI_ISL_285605 | EPI1104214 | HA | UK | 2017-Nov-22 | A/Singapore/INFIMH-16-0019/2016 IVR-186 (17/210) | National Institute for Biological Standards and Control | National Institute for Biological Standards and Control |
| EPI_ISL_282213 | EPI1082230 | HA | UK | 2017-Oct-16 | A/Singapore/INFIMH-16-0019/2016 NIB-104 (17/194) | National Institute for Biological Standards and Control | National Institute for Biological Standards and Control |
| EPI_ISL_293533 | EPI1151848 | HA | USA | 2016 | A/Singapore/Infimh-16-0019/2016 X-307 | New York Medical College | Centers for Disease Control and Prevention |
| EPI_ISL_293527 | EPI1151800 | HA | USA | 2016 | A/Singapore/Infimh-16-0019/2016 X-307A | New York Medical College | Centers for Disease Control and Prevention |

**Supplementary Table 3**

Historical occurrences of egg-adaptation issues with A(H3N2) vaccines.  Several times the WHO has chosen a clinical specimen reference strain for the CVVs that is well-matched to the dominant circulating virus strain during that season, but the egg-passaged CVV strain (or "like" strain) adapted with critical antigenic changes.  We quantify the antigenic distance between the egg-passaged CVV and the dominant circulating virus from each of these seasons using $p_{epitope}$.  In the years listed here, $p_{epitope}$ between the reference strain and the dominant circulating strain is zero.  Accession numbers below each strain are from GISAID [1].

| Season | Candidate Vaccine Virus Reference (Clinical Specimen) | Candidate Vaccine Virus (Egg passaged) | Dominant Circulating Strain | Dominant epitope | $p_{epitope}$ | Observed Vaccine Efficacy |
|---|---|---|---|---|---|---|
| 1996-97 | A/Wuhan/359/1995 EPI_ISL_758 - EPI3775 | A/Nanchang/933/1995 EPI_ISL_167267 - EPI545305 | A/Wuhan/359/1995 EPI_ISL_758 - EPI3775 | B | 0.0952 | 28% [11] |
| 2004-05 | A/Fujian/411/2002 EPI_ISL_111384 - EPI362915 | A/Wyoming/3/2003 EPI_ISL_3767 - EPI160210 | A/Fujian/411/2002 EPI_ISL_111384 - EPI362915 | B | 0.0952 | 9% [12] |
| 2011-12 | A/Victoria/361/2011 EPI_ISL_167307 - EPI545346 | A/Perth/16/2009 EPI_ISL_176456 - EPI577972 | A/Victoria/361/2011 EPI_ISL_167307 - EPI545346 | C | 0.1111 | 23% [12] |
| 2012-13 | A/Victoria/361/2011 EPI_ISL_167307 - EPI545346 | A/Victoria/361/2011 EPI_ISL_132936 - EPI408194 | A/Victoria/361/2011 EPI_ISL_167307 - EPI545346 | B | 0.0952 | 35% [12] |

vaccine for the Northern Hemisphere 2017-2018.  Available at https://www.crick.ac.uk.  Accessed 1 February 2018.